\documentclass{article}

\usepackage{lmodern}
\usepackage[a4paper, total={6.5in,9in}]{geometry}
\usepackage{graphicx} 
\usepackage{authblk}
\usepackage{subcaption} 
\usepackage{wrapfig}
\usepackage{svg}
\usepackage[most]{tcolorbox}
\usepackage[table]{xcolor}
\usepackage{colortbl}
\usepackage{booktabs}
\usepackage{mathtools}
\usepackage{amsmath, amssymb, amsfonts,amsthm}

\usepackage{hyperref}
\usepackage{tabularx}
\usepackage{array} 
\usepackage{float}
\usepackage{multirow}
\usepackage{tikz}
\usetikzlibrary{positioning, matrix, arrows, calc, arrows.meta, shapes, fit}
\usepackage[skip=4pt]{caption}

\usepackage{titlesec}
\titlespacing{\section}{0pt}{6pt}{4pt}
\titlespacing{\subsection}{0pt}{4pt}{2pt}

\author{Adeela Bashir$^1$}
\author{Zia Ush Shamszaman$^{2}$}
\author{Zhao Song$^1$}
\author{The Anh Han$^{1,3,*}$}
\affil{$^{1}$School of Computing, Engineering and Digital Technologies, Teesside University\\ $^{2}$Department of Informatics, Faculty of Natural, Mathematical \& Engineering Sciences, King's College London: London, UK \\ $^{3}$Center for Digital Innovation, Teesside University\\ $^*$Corresponding: The Anh Han (T.Han@tees.ac.uk)}

\begin{document}

\title{Network Reciprocity Shapes Evolutionary Cybersecurity Dynamics} 

\date{\today}
\maketitle

\begin{abstract}

AI-assisted cybersecurity systems are characterised by continuous adaptation between attackers and defenders, making evolutionary game theory a natural framework for studying their long-term behaviour. However, existing evolutionary cybersecurity models have primarily focused on homogeneous interactions, providing limited understanding of how population structure influences cyber attack--defence dynamics. In this paper, we develop a mixed-role evolutionary game in which adaptive cyber agents can exhibit both offensive and defensive behaviours, and investigate its dynamics in well-mixed and structured populations. The proposed framework combines stochastic evolutionary analysis with large-scale agent-based simulations to examine how interaction structure shapes long-run strategic behaviour. Our results reveal a fundamental difference between global and local interactions. While well-mixed populations exhibit broad coexistence between attacking and defensive strategies, structured populations generate well-defined evolutionary regimes in which secure defensive behaviour becomes dominant over a much larger region of the parameter space. Spatial analysis further shows that neighbouring defenders naturally form resilient clusters that suppress persistent attacks through network reciprocity. These findings demonstrate that interaction structure is a fundamental determinant of AI-assisted cybersecurity evolution and suggest that organising defensive agents through local networked interactions can substantially improve long-term cyber resilience without requiring additional defensive incentives.
\end{abstract}

\section{Introduction}

Artificial intelligence (AI) is rapidly transforming modern cybersecurity. Large language models (LLMs), autonomous agents, and machine learning techniques are increasingly being used by both attackers and defenders. AI enables attackers to automate phishing campaigns, generate malicious code, launch adaptive malware, and exploit vulnerabilities at unprecedented speed. At the same time, AI-assisted defence improves intrusion detection, malware analysis, behavioural analytics, threat intelligence, and automated incident response, enabling organisations to respond more rapidly to evolving threats \cite{akinyemi2024ai,guembe2022emerging,saeri2026prioritization}. Thus, cybersecurity has become an adaptive evolutionary process in which both sides co-evolve by continuously modifying their strategies in response to changing incentives and environmental conditions \cite{bashir2026co}.

Evolutionary Game Theory (EGT) provides a natural framework for analysing such adaptive cybersecurity dynamics \cite{dorsey2003mathematical,verma2024exploring,tirenin1999concept,hofbauer1998evolutionary}. Unlike classical game theory, EGT captures how competing behaviours spread, persist, and evolve over time, making it particularly suitable for studying long-term attack-defence interactions \cite{smith1973logic,sigmund2010calculus}. Consequently, EGT has been widely applied to cybersecurity investment, intrusion response, attack-defence games, and collective defence mechanisms \cite{skopik2022detecting,do2017game,hilbe2023evolutionary}. These studies have improved our understanding of strategic cyber conflicts by revealing how incentives influence the evolutionary success of attacking and defensive behaviours \cite{bashir2026strategic}.

However, most existing evolutionary cybersecurity models focus primarily on strategic incentives \cite{reittinger2026share,wessels2021understanding}. In many real-world cyber systems, organisations, devices, and AI-enabled security agents interact through communication networks, enterprise infrastructures, cloud platforms, and supply-chain relationships rather than through homogeneous random mixing \cite{melnyk2022new,odimarha2024securing,boiko2019information}. Extensive studies in EGT have demonstrated that interaction structure can fundamentally reshape evolutionary outcomes by altering strategy propagation, invasion dynamics, and equilibrium selection in networked populations \cite{Szolnoki2013,nathanson2009calculating,szabo2007evolutionary,cimpeanu2022artificial,pi2026evolutionary}. However, these insights have rarely been translated to cybersecurity. Existing EGT models explain how incentives influence attacker–defender behaviour, but cannot explain how attacks propagate through connected cyber infrastructures or how local defensive coordination emerges. It remains unclear how population structure influences the long-term evolution of attack-defence dynamics and the emergence of cyber resilience in interconnected systems.

\begin{figure}[t]
\centering
    \includegraphics[width=0.85\textwidth]{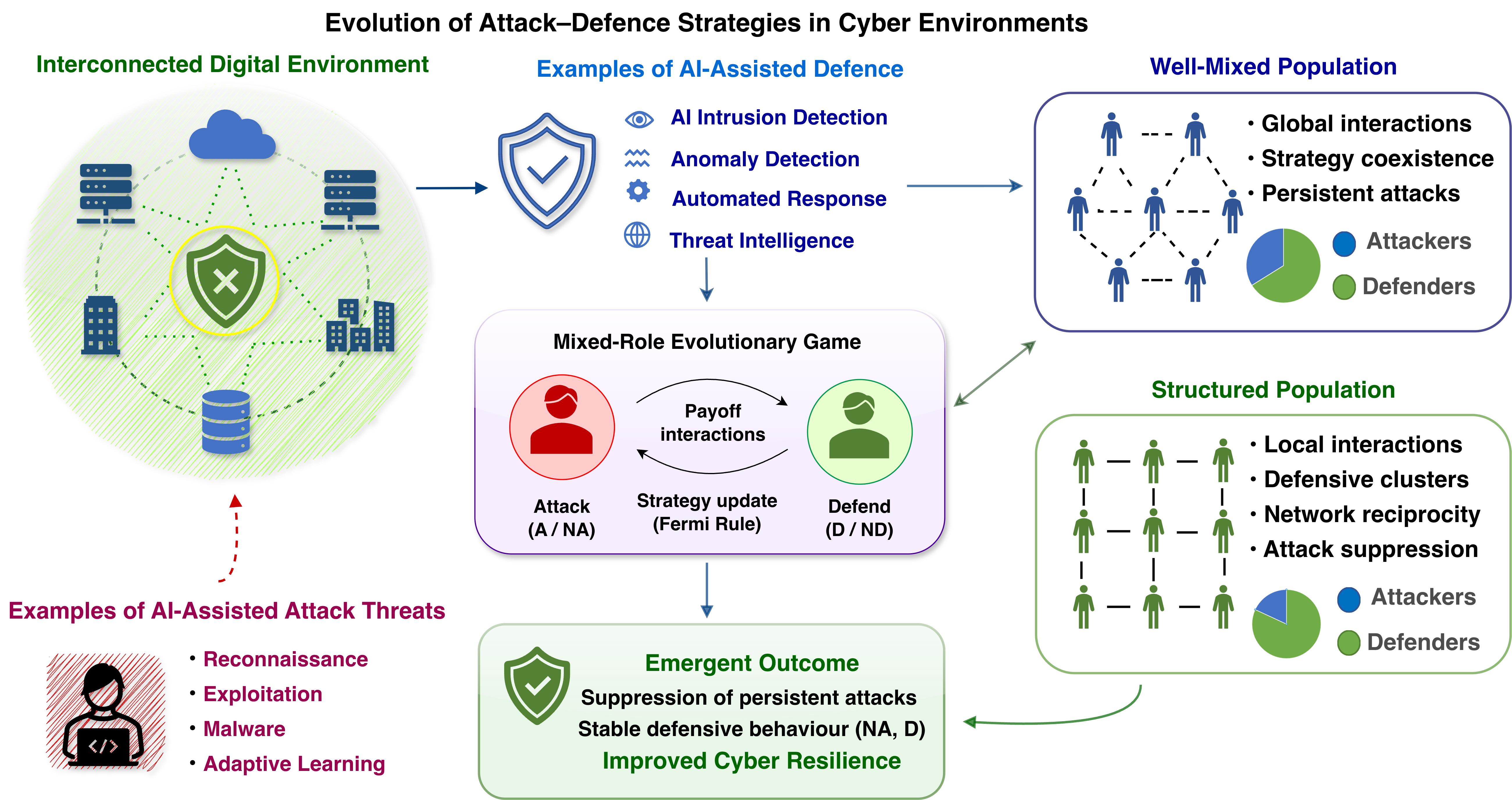}
\caption{
Conceptual overview of the proposed framework where attacks and defence mechanisms interact within a  cyber environment and interactions are modelled using a mixed-role evolutionary game.
}
\label{fig:1}
\end{figure}

Motivated by this gap, we propose an evolutionary game-theoretic framework to investigate how population structure influences the long-run evolution of adaptive AI-assisted cyber attack-defence dynamics. The overall workflow of the proposed framework is illustrated in Figure~\ref{fig:1}. AI-assisted cyber attacks and AI-enabled defence mechanisms interact within a secure digital environment and are represented using a mixed-role evolutionary game. The resulting evolutionary dynamics are analysed under both well-mixed and structured populations to quantify the impact of interactions on attack suppression, defensive clustering, and long-term cyber resilience. 

Our results demonstrate that population structure fundamentally influences the evolutionary dynamics of cyber conflict. While well-mixed populations exhibit broad coexistence between attacking and defensive behaviours, structured populations promote defensive clustering through network reciprocity, suppress persistent attacks, and drive the system towards secure defensive states. These findings provide new insights into how network organisation can enhance cyber resilience alongside advances in AI-assisted defence. 

The main contributions of this work are:

\begin{itemize}
    \item We develop an evolutionary game model to study AI-assisted cyber attack-defence dynamics in both well-mixed and structured populations.

    \item We combine stochastic evolutionary analysis with large-scale agent-based simulations to compare the long-run behaviour of homogeneous and networked cyber systems.

    \item We show that network reciprocity promotes defensive clustering and suppresses persistent attacks.

    \item We provide analytical and computational evidence that interaction structure is a key factor governing the evolution of AI-assisted cyber resilience.
\end{itemize}

The remainder of this paper is organised as follows. Section~\ref{sec:related} reviews the related work on EGT and AI-assisted cybersecurity. Section~\ref{sec:model} presents the proposed AI-assisted cyber attack-defence game, derives the attack-suppression threshold, and formulates the evolutionary dynamics in both well-mixed and structured populations. Section~\ref{sec:results} presents the simulation setup and discusses the results, including stochastic evolutionary dynamics, the comparison between well-mixed and structured populations, with particular emphasis on defensive clustering as a mechanism for attack suppression. Finally, Section~\ref{sec:conclusion} concludes the paper and outlines directions for future research.

\section{Related Work}
\label{sec:related}

The rapid adoption of AI has transformed both offensive and defensive cybersecurity capabilities. AI-assisted techniques are increasingly used for vulnerability discovery, malware generation, phishing campaigns, intrusion detection, malware analysis, and automated incident response. As AI systems become more autonomous, attackers and defenders continuously adapt their behaviour in response to one another, creating highly dynamic cyber environments \cite{petihakis2024aias,bashir2026strategic,skopik2022detecting}. These developments have motivated growing interest in mathematical models that can capture the long-term evolution of AI-assisted cyber conflicts rather than analysing isolated attack or defence decisions.

Evolutionary game-theoretic models have been widely applied to study adaptive cybersecurity problems, including cybersecurity investment, attack-defence interactions, intrusion response, malware propagation, trust management, and collaborative defence \cite{do2017game,khalid2023recent,verma2024exploring,tirenin1999concept}.Existing studies consistently demonstrate that the long-term success of attackers and defenders depends on the balance between attack incentives, defensive costs, and protection effectiveness \cite{bashir2026co}. More recent work has further incorporated adaptive adversaries and AI-assisted defence, providing valuable insights into how intelligent cyber agents continuously modify their strategies in response to changing threat environments \cite{petihakis2024aias,bashir2026strategic}. Complementing these evolutionary approaches, broader game-theoretic cybersecurity models have investigated collaborative intrusion detection, industrial cyber-physical system security, advanced persistent threats, blockchain security, and resilient control, primarily focusing on application-specific defence optimisation and equilibrium analysis rather than evolutionary population dynamics \cite{wu2018game,huang2019game,zhang2022game,feng2018cyber,liu2026game}. However, these studies predominantly analyse strategic adaptation under homogeneous interaction settings, where every agent is assumed to interact equally with every other agent.

In contrast, interaction structure has been extensively studied in EGT \cite{Szolnoki2013,nathanson2009calculating,song2026network}. Research on cooperation and social dilemmas has shown that network topology fundamentally alters evolutionary outcomes by enabling cooperative agents to form resilient local clusters through network reciprocity \cite{szabo1998evolutionary,nowak2006five,perc2017statistical,flores2022cooperation,Han2018fostering}. Structured populations exhibit spatial self-organisation, phase transitions, and stable behavioural patterns that do not emerge under homogeneous mixing \cite{song2026network,hofbauer1998evolutionary,duh2020mixing,santos2006evolutionary}. These findings suggest that interaction structure may play an equally important role in cybersecurity, where attacks propagate through communication networks, enterprise infrastructures, cloud systems, and interconnected organisations rather than through globally mixed populations \cite{melnyk2022new,odimarha2024securing,boiko2019information}.

Despite these advances, the role of population structure in AI-assisted cybersecurity remains largely unexplored. Understanding how cyber behaviours propagate through interconnected populations, how local interactions reshape attack persistence and defensive organisation, and whether network organisation can suppress persistent attacks and promote cyber resilience is therefore an important open research question. To address this gap, we develop an evolutionary game model for AI-assisted cyber attack-defence interactions and investigate its dynamics in both well-mixed and structured populations. By integrating stochastic evolutionary analysis with large-scale agent-based simulations, we systematically compare homogeneous and local interaction structures and demonstrate how network reciprocity suppresses attacks, promotes defensive clustering, and expands the parameter region in which secure defensive behaviour emerges.

\section{AI-Assisted Cyber Attack-Defence Game}
\label{sec:model}

Guided by the conceptual framework shown in Figure~\ref{fig:1}, we now formalise the interactions between attackers and defenders as an evolutionary game. We first define the strategy space and payoff structure before analysing the resulting evolutionary dynamics in well-mixed and structured populations. The payoffs associated with the attacker and defender actions are summarised in Table~\ref{tab:1}, and AI-assisted capabilities are represented through the effectiveness of attack and defence. The attacker incurs an attack cost $c_a$ for launching an attack against an undefended target and receives the attack benefit $b_a$, while the defender suffers a loss equal to the asset value $w$. If defence is deployed, the defender additionally incurs the defence cost $c_d$, whereas successful defence prevents the attack with probability $p_d$, reducing the attacker's expected benefit while increasing the defender's protection.  In particular, $p_d$ represents the effectiveness of AI-assisted defensive mechanisms, such as intelligent intrusion detection, automated response, and AI-driven threat mitigation.

\begin{table}[H]
   \centering
   \caption{Payoffs of the adaptive attacker and defender}
    \begin{tabular}{|c|c|c|c|}
      \hline
      \multicolumn{2}{|c|}{\textbf{Strategies}} & \multicolumn{2}{|c|}{\textbf{Payoffs}} \\
      \hline
     \textbf{D} & \textbf{A} & \textbf{Defender} & \textbf{Attacker}\\
     \hline
     $ND$ & $NA$ & 0 & 0 \\
     \hline
     $ND$ & $A$ & $-w$ & $-c_a+b_a$ \\
     \hline
     $D$ & $NA$ & $-c_d+b_d$ & 0 \\
     \hline
     $D$ & $A$ & $-c_d+p_db_d-w(1-p_d)$ & $-c_a+b_a(1-p_d)$ \\
     \hline
  \end{tabular}
    \label{tab:1}
\end{table}
The model contains six parameters describing the costs, benefits, and effectiveness of cybersecurity actions. Their meanings are summarised in Table~\ref{tab:2}.
\begin{table}[H]
    \centering
    \renewcommand{\arraystretch}{1.5}
        \caption{Summary of parameters in the model}
    \label{tab:2}
    \begin{tabular}{|c|m{8cm}|c|}
    \hline
    \textbf{Variables} & \textbf{Meaning of variables} & \textbf{Constraints}\\
    \hline
    $w$  & Assets value, i.e. loss to the defender for an attack & $0 < w \leq1$ \\
     \hline
     $c_a$  & Cost to the attacker for attack attempt & $ 0< c_a < w $ \\
    \hline
    $c_d$  & Cost to the defender for implementing defence & $ 0 < c_d <w $ \\
    \hline
     $b_a$  & Attacker's benefit for a successful attack & $c_a < b_a$ \\
     \hline
    $b_d$  & Defender’s benefit for not being breached & $c_d < b_d \leq w$ \\
     \hline
     $p_d$  & Probability of successful defence & $0 < p_d\leq1$ \\
     \hline
    \end{tabular}
\end{table}

Each agent can choose among two strategies. Attacking behaviour $AT\in\{A,NA\}$, where A denotes launching an AI-assisted cyber attack, and NA denotes not attacking. Similarly, defensive behaviour $DF\in\{D,ND\}$, where D denotes deploying AI-assisted defence, and ND denotes not investing in defensive protection. Combining the offensive and defensive decisions produces four mixed-role strategies, $(A,D),\;
(A,ND),\;
(NA,D),\;
(NA,ND).$

These four mixed-role strategies represent distinct cybersecurity behaviours:

\begin{itemize}
    \item $(A,D)$ (\textit{active defence}), where organisations actively attack others while simultaneously investing in protecting themselves;
    
    \item $(NA,D)$ (\textit{defensive-only}), where organisations invest in self-defence and do not attack others;
    
    \item $(A,ND)$ (\textit{offensive-only}), where organisations prioritise offensive capability but neglect self-protection;
    
    \item $(NA,ND)$ (\textit{passive}), where organisations neither launch attacks nor invest in defensive mechanisms.
\end{itemize}

These four strategies capture the principal behavioural modes observed in modern AI-assisted cybersecurity environments, ranging from proactive defence to purely offensive and passive security postures. Modern cybersecurity organisations increasingly possess both offensive and defensive capabilities \cite{huma2025hacking}. National cyber agencies and military cyber units conduct defensive operations while also maintaining offensive cyber capabilities \cite{saltzman2013cyber}. Likewise, organisations routinely employ internal red teams that perform penetration testing and adversarial assessments alongside blue teams responsible for protecting operational infrastructure \cite{kumar2024ai}. Commercial cybersecurity vendors similarly combine offensive security testing with defensive monitoring and incident response services. These examples motivate representing cyber agents as mixed-role entities capable of simultaneously adopting offensive and defensive behaviours.

In the proposed model, during every interaction, one player first assumes the attacker role while the opponent acts as the defender. The roles are then exchanged, allowing each player to experience both offensive and defensive interactions. The overall payoff is defined as the average utility obtained across the two roles. Formally, the payoff of a player with strategy $(AT_1, DF_1)$ against a player with strategy $(AT_2, DF_2)$ is given by:
\[
\pi\big((AT_1, DF_1),(AT_2, DF_2)\big)
=
\frac{1}{2}\Big[
\pi(AT_1, DF_2)
+
\pi(DF_1, AT_2)
\Big],
\]
where $\pi(AT_i,DF_j)$ denotes the payoff obtained when the focal player performs the attacking role against the opponent's defensive strategy, and $\pi(DF_i,AT_j)$ denotes the payoff obtained when acting as the defender against the opponent's attacking strategy. Averaging over the two interaction roles reflects the dual offensive and defensive capabilities of modern AI-assisted cyber agents.

The resulting mixed-role payoff matrix is given in Table~\ref{tab:3}. The payoff structure determines the evolutionary incentives governing cyber attack and defence.
\begin{table}[H]
\centering
\renewcommand{\arraystretch}{1.5}
\caption{Payoff matrix for mixed-role game.}
\begin{tabular}{c|cccc}
 & $(A,D)$ & $(A,ND)$ & $(NA,D)$ & $(NA,ND)$ \\
\hline
$(A,D)$ & $\frac{1}{2}(\pi_A^D + \pi_D^A)$ 
         & $\frac{1}{2}(\pi_A^{ND} + \pi_{ND}^A)$ 
         & $\frac{1}{2}(\pi_A^D + \pi_D^{NA})$ 
         & $\frac{1}{2}(\pi_A^{ND})$ \\

$(A,ND)$ & $\frac{1}{2}(\pi_A^D + \pi_D^A)$ 
          & $\frac{1}{2}(\pi_A^{ND} + \pi_{ND}^A)$ 
          & $\frac{1}{2}(\pi_A^D + \pi_D^{NA})$ 
          & $\frac{1}{2}(\pi_A^{ND})$ \\

$(NA,D)$ & $\frac{1}{2}(\pi_D^A)$ 
          & $\frac{1}{2}(\pi_{ND}^A)$ 
          & $\frac{1}{2}(\pi_D^{NA})$ 
          & $0$ \\

$(NA,ND)$ & $\frac{1}{2}(\pi_D^A)$ 
           & $\frac{1}{2}(\pi_{ND}^A)$ 
           & $\frac{1}{2}(\pi_D^{NA})$ 
           & $0$ \\
\end{tabular}
\label{tab:3}
\end{table}


\paragraph{Analytical Condition for Attack Suppression.}
The payoff structure of the proposed game provides a simple analytical condition that explains when attacking ceases to be evolutionarily profitable. Consider an interaction in which an attacking agent encounters a defended opponent. From Table~\ref{tab:1}, the expected payoff of the attacker is

\[
\pi_A^D=-c_a+b_a(1-p_d),
\]

where $c_a$ is the operational cost of launching an attack, $b_a$ is the benefit of a successful attack, and $p_d$ denotes the probability that the defender successfully detects and blocks the attack.

For an attacking strategy to remain evolutionarily favourable, its expected payoff must be positive, i.e.,

\[
\pi_A^D>0.
\]

Substituting the attacker payoff yields

\[
-c_a+b_a(1-p_d)>0,
\]

which can be rearranged to show attacking strategy is evolutionarily advantageous against defended agents if and only if

\[
b_a(1-p_d)>c_a.
\label{eq:attack_threshold}
\]

Equivalently, attacks become evolutionarily unattractive whenever

\begin{equation}
p_d>1-\frac{c_a}{b_a}.
\label{eq:attack_condition}
\end{equation}

Equation~(\ref{eq:attack_condition}) provides an analytical attack-suppression condition. When the probability of successful defence exceeds this threshold, the expected payoff of attacking becomes non-positive, making attacks evolutionarily unattractive against defended agents. It shows that attack suppression is governed jointly by three factors: the operational cost of launching attacks ($c_a$), the reward obtained from successful attacks ($b_a$), and the effectiveness of AI-assisted defensive mechanisms ($p_d$). Increasing the operational cost of attacks or improving defence both reduce the evolutionary attractiveness of offensive behaviour. Although this condition follows directly from the payoff structure, it provides useful analytical intuition for interpreting the phase transitions and attack-suppression behaviour observed in the stochastic analysis and structured population simulations presented in the following sections.

\subsection{Well-Mixed Finite Population}

We consider a well-mixed finite population consisting of $M$ players, where every player can interact with evey other player, as shown in Figure~\ref{fig:1}. Each player adopts a strategy $S$ from the four mixed-role strategies where $\mathcal{S} = \{(A,D), (A,ND), (NA,D), (NA,ND)\}.$ At each time step, a randomly selected player updates its strategy by imitating the strategy of another randomly selected player, following a Moran-type process \cite{traulsen2006stochastic,bashir2026strategic}. To analyse the evolutionary dynamics, we consider pairwise competition between two strategies $X, Y \in \mathcal{S}$. Suppose there are $m$ players adopting strategy $X$ and $M-m$ players adopting strategy $Y$. The average payoffs of players using strategies $X$ and $Y$ are given by:
\[
f_X = \frac{(m - 1)\pi_{X,X} + (M - m)\pi_{X,Y}}{M - 1}, \quad
f_Y = \frac{m\pi_{Y,X} + (M - m - 1)\pi_{Y,Y}}{M - 1},
\]
where $\pi_{X,Y}$ denotes the payoff of a player using strategy $X$ when interacting with a player using strategy $Y$, as defined by the mixed-role payoff matrix in Table~\ref{tab:3}. The evolutionary dynamics are governed by the Fermi update rule \cite{traulsen2006stochastic}, where a player with strategy $X$ randomly selects one of its neighbours and adopts their strategy $Y$ with probability:
\begin{equation}
    P(X \rightarrow Y) = \frac{1}{1 + \exp[\beta(f_Y - f_X)]},
    \label{eq:2}
\end{equation}
where $\beta$ represents the selection intensity, controlling how strongly payoff differences influence strategy adoption. When $\beta \rightarrow 0$, strategy updates are random (neutral drift), while for large $\beta$, players are more likely to imitate strategies with higher payoffs. Based on this update rule, the probability that the number of players using strategy $X$ increases or decreases by one is given by:
\[
T^{\pm}_{XY} = \frac{M - m}{M}  \frac{m}{M}  
\left[1 + \exp(\mp \beta(f_X - f_Y))\right]^{-1}.
\]

\noindent The fixation probability of a single mutant with strategy $X$ in a resident population of $M-1$ players using strategy $Y$ is given by:
\[
\rho_{YX} =
\left[
1 + \sum_{k=0}^{M-1} \prod_{m=1}^{k}
\frac{T^{-}_{XY}}{T^{+}_{YX}}
\right]^{-1}.
\]

\noindent Assuming a small mutation limit, where each mutant either fixates or goes extinct before another mutation occurs, the population transitions only between homogeneous strategic states. This assumption reflects the fact that the introduction of fundamentally new AI-assisted attack or defence strategies is much less frequent than routine cyber interactions, while also making the stochastic evolutionary dynamics analytically tractable. The fixation probabilities $\rho_{XY}$ therefore define the transition probabilities between homogeneous states of the population. Let $q=4$ be the number of strategies. The transition matrix is defined as:
\[
T_{XY} =
\begin{cases}
\frac{\rho_{XY}}{q-1}, & X \neq Y, \\
1 - \sum_{{Y=1},{Y \neq X}}^q T_{XY}, & X = Y.
\end{cases}
\]

The stationary distribution of this Markov process is obtained from the normalised eigenvector corresponding to eigenvalue 1 of the transposed transition matrix \cite{sigmund2010calculus,traulsen2006stochastic}. This distribution describes the long-run proportion of time the population spends in each homogeneous state, providing a measure of the evolutionary success of each mixed-role strategy.
\begin{figure*}[t!]
\centering

\subfloat[]{%
    \includegraphics[width=0.25\textwidth]{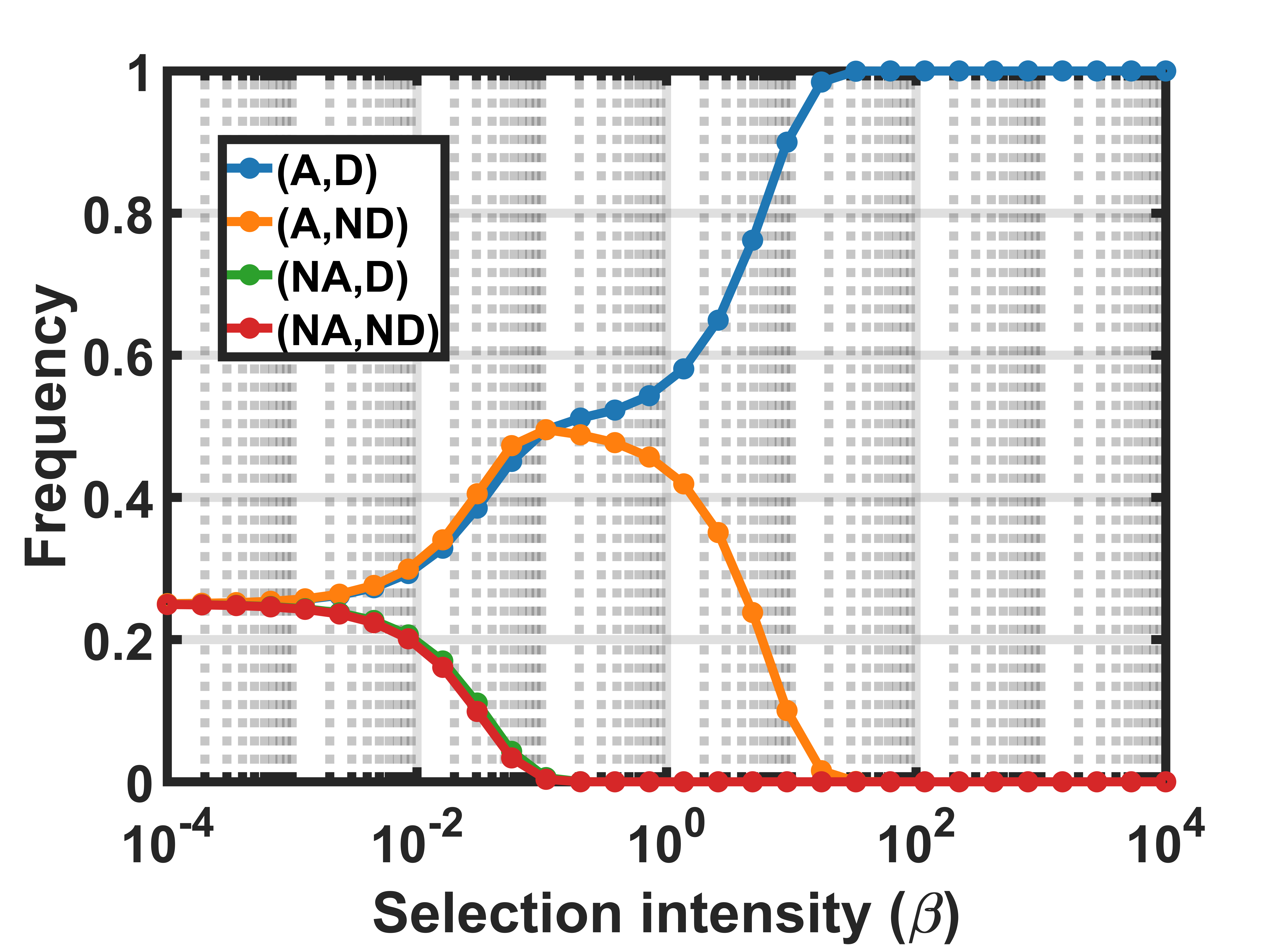}%
    \label{fig:2a}%
}
\hspace{-1em}
\subfloat[]{%
    \includegraphics[width=0.25\textwidth]{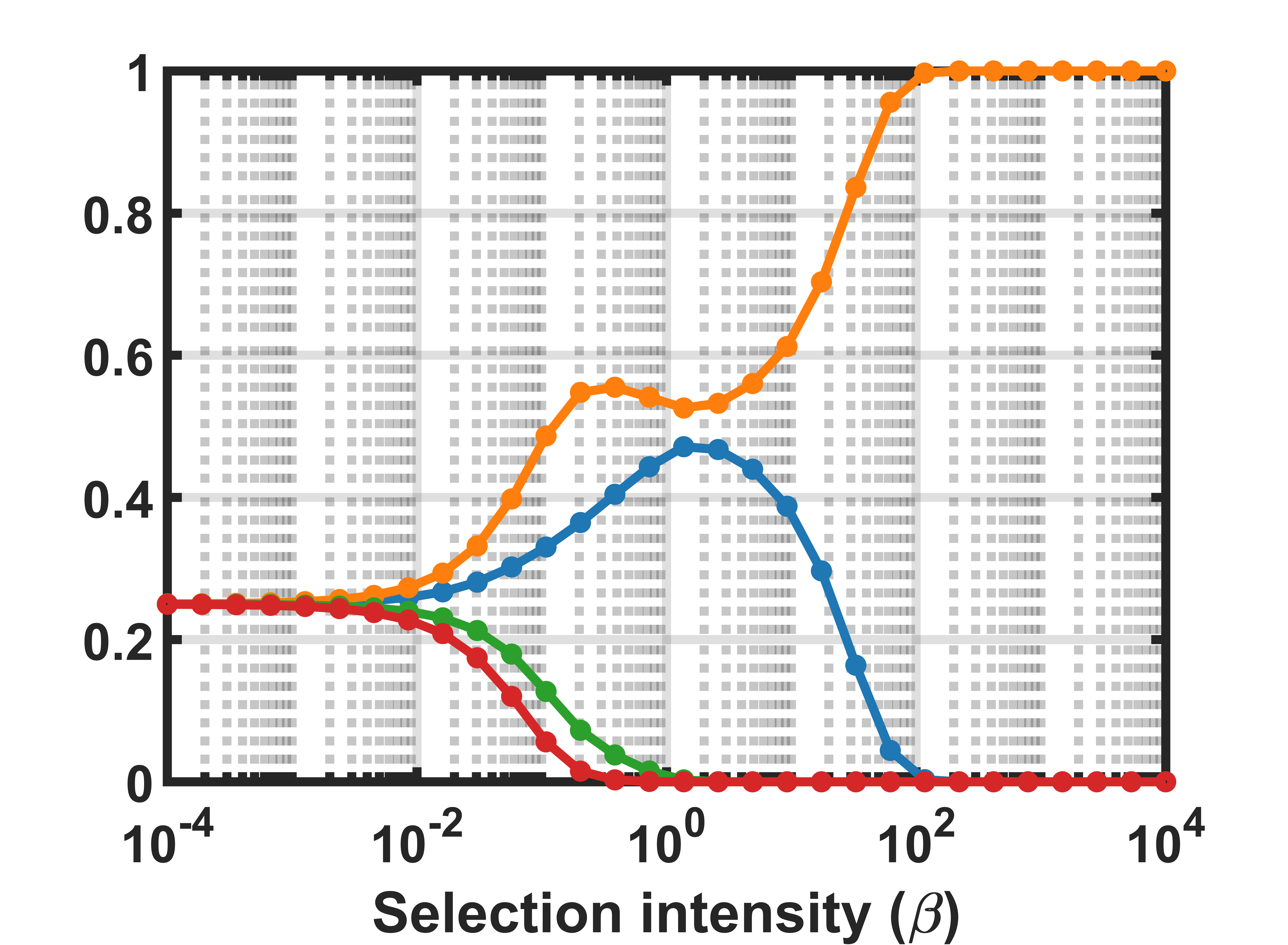}%
    \label{fig:2b}%
}
\hspace{-1em}
\subfloat[]{%
    \includegraphics[width=0.25\textwidth]{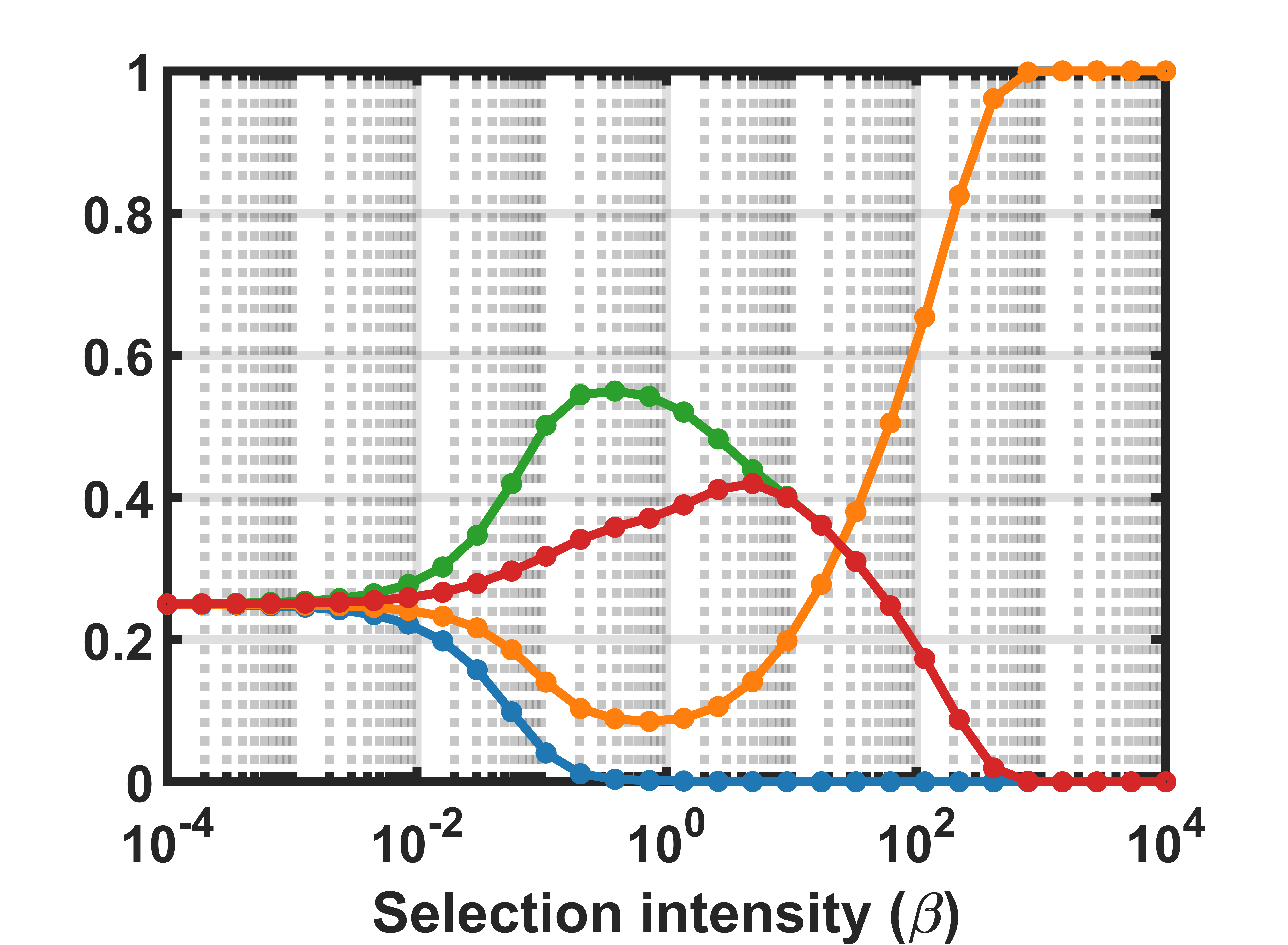}%
    \label{fig:2c}%
}
\hspace{-1em}
\subfloat[]{%
    \includegraphics[width=0.25\textwidth]{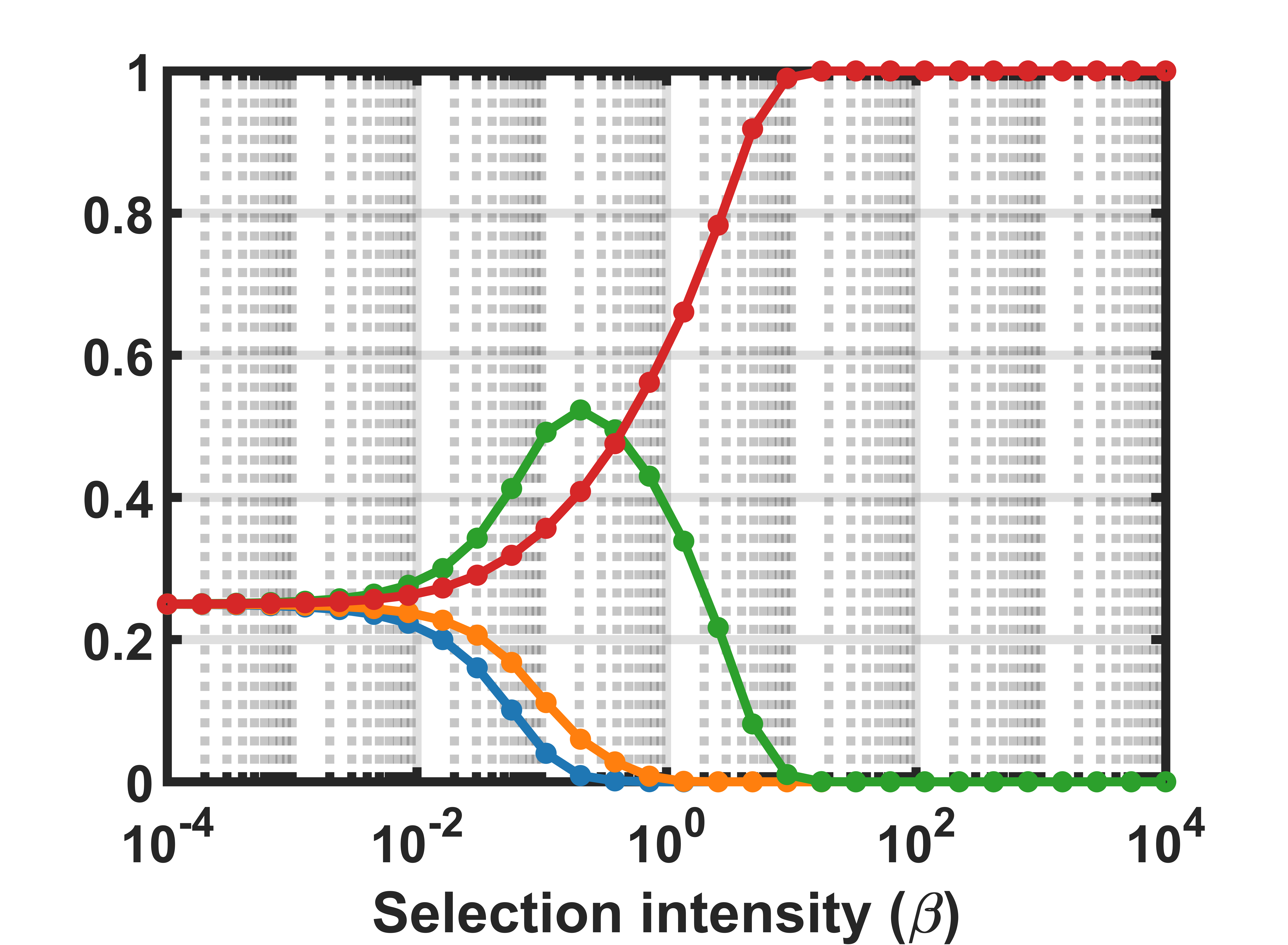}%
    \label{fig:2d}%
}

\caption{Stationary frequencies of AI-assisted cyber attack-defence strategies as functions of the selection intensity $\beta$ under four representative cybersecurity environments. Weak selection is dominated by random drift, whereas stronger selection increasingly favours strategies with higher payoffs. Panels (a) and (b) represent attack-favourable environments in which AI-assisted offensive strategies dominate. Panels (c) and (d) illustrate defence-favourable environments, where higher attack costs and more effective AI-assisted defence progressively suppress attacks and promote non-attacking strategies. Parameter values for (a) $c_a = 0.2, b_a = 1.0, p_d = 0.2,
c_d = 0.5, b_d = 1.0, w = 1.0$, (b) $c_a = 0.45,b_a = 1.0,p_d = 0.5,c_d = 0.4,b_d = 1.0,w= 1.0$, (c) $c_a = 0.8,b_a = 0.8,p_d = 0.9,c_d = 0.2,b_d = 1.2,w = 1.0$, and (d) $c_a = 0.8,b_a = 0.7,p_d = 0.8;,c_d = 0.2,b_d = 1.2,w = 1.0$.}
\label{fig:2}
\end{figure*}
\subsection{Structured Population}

To investigate the influence of interaction structure, we next consider a structured population represented by a two-dimensional square lattice with periodic boundary conditions \cite{szabo2007evolutionary}. Each node corresponds to one adaptive cyber agent, and each agent interacts only with its four nearest neighbours, as shown in Figure~\ref{fig:1}. This local interaction pattern reflects the fact that cyber attacks and defensive responses often propagate through connected infrastructures rather than through globally mixed populations. During each Monte Carlo Step (MCS), a randomly selected agent compares its accumulated payoff with one randomly chosen neighbour and updates its strategy according to the Fermi rule in Eq.~(\ref{eq:2}). The process is repeated until the population reaches a stationary state, from which the long-run frequencies of the four strategies are obtained. Further details on the simulations setup are given in Section 4.3.

\section{Results and Discussion}
\label{sec:results}

In this section, we investigate the evolutionary dynamics of the proposed AI-assisted cyber attack-defence game using both analytical and computational approaches. We first analyse the stochastic evolutionary dynamics of the finite well-mixed population through the Markov-chain framework, followed by a parameter sensitivity analysis to identify the key factors governing attack and defence. We then compare the evolutionary behaviour of well-mixed and structured populations using agent-based simulations and finally examine how defensive clustering suppresses attacks and promotes cyber resilience.

\subsection{Stochastic Evolutionary Dynamics}

We first investigate how selection pressure influences the long-run evolution of AI-assisted cyber attack-defence strategies in a well-mixed population. The stochastic Markov framework enables us to identify the stationary behaviour of the four mixed-role strategies under different cybersecurity conditions, providing a baseline for comparison with the structured population model introduced later. Figure~\ref{fig:2} illustrates the stationary frequencies of the four AI-assisted cyber strategies as the selection intensity $\beta$ increases under different cybersecurity environments. When selection is weak $(\beta<10^{-2})$, all strategies occur with similar frequencies because strategy updates are dominated by random drift rather than payoff differences. As selection becomes stronger, evolutionary success is increasingly determined by the balance between attack incentives and defensive effectiveness.

\begin{figure}[b!]
\centering

\begin{subfigure}[t]{0.24\textwidth}
\centering
\includegraphics[width=\linewidth]{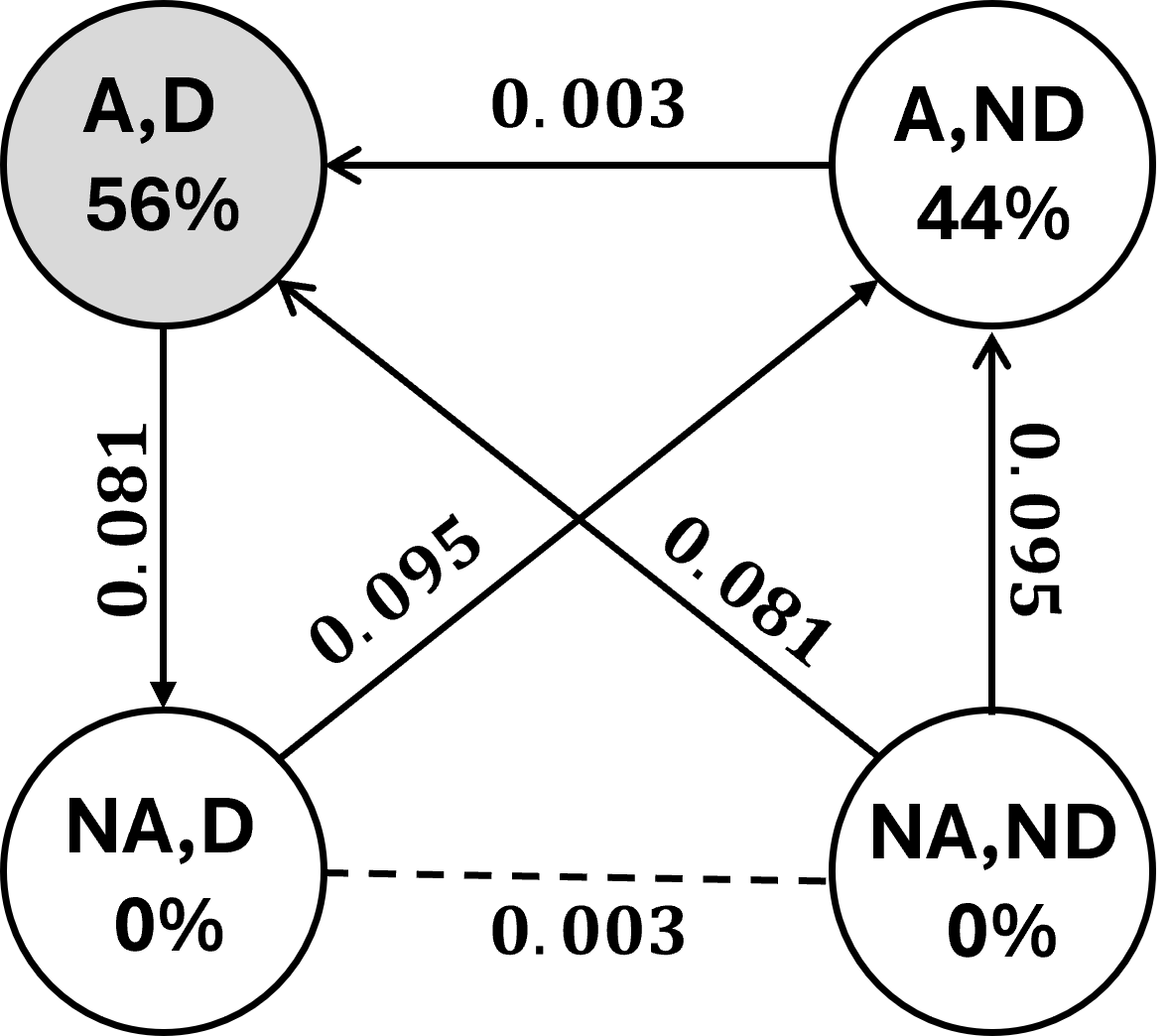}
\caption{}
\end{subfigure}
\hfill
\begin{subfigure}[t]{0.24\textwidth}
\centering
\includegraphics[width=\linewidth]{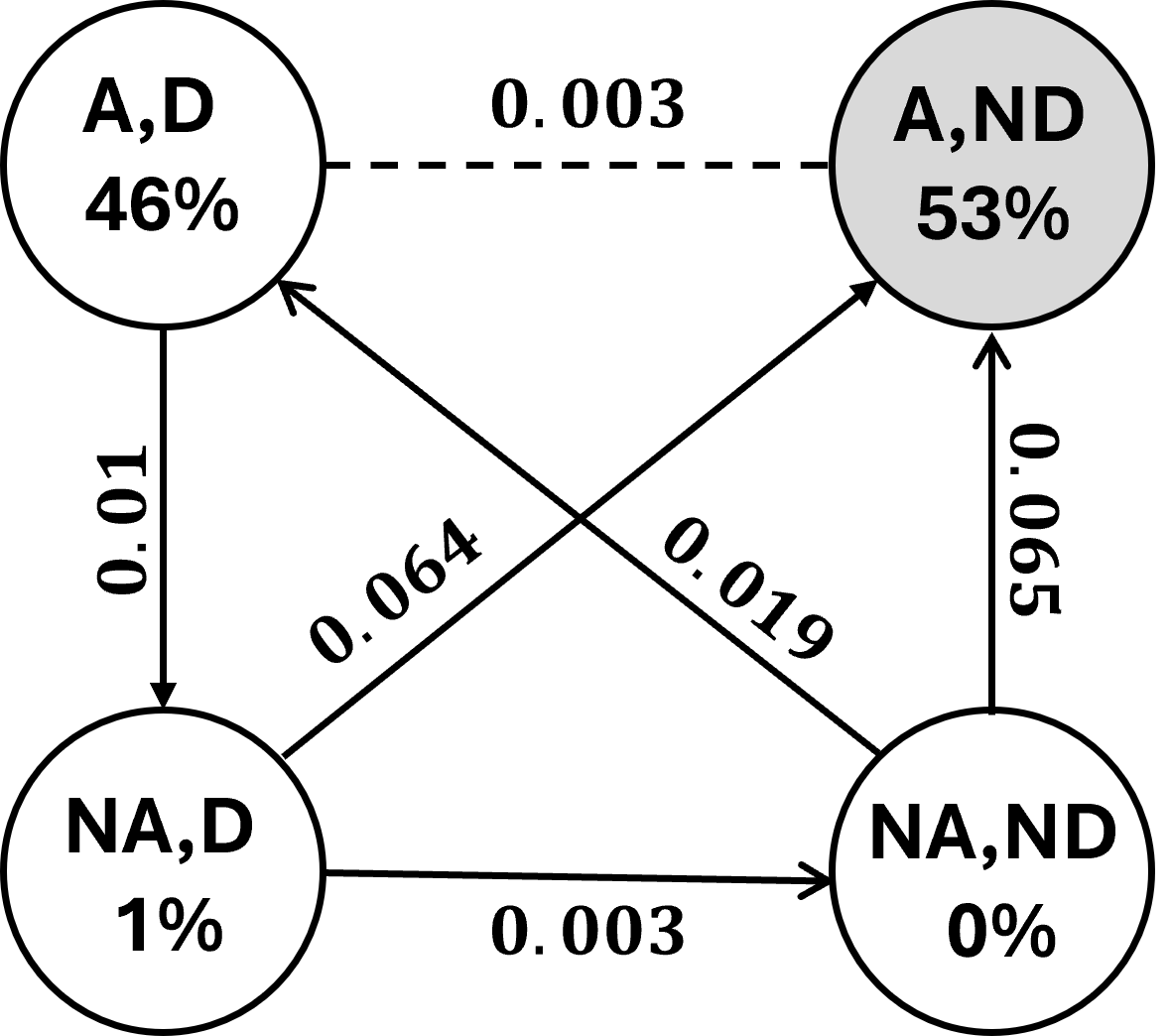}
\caption{}
\end{subfigure}
\hfill
\begin{subfigure}[t]{0.24\textwidth}
\centering
\includegraphics[width=\linewidth]{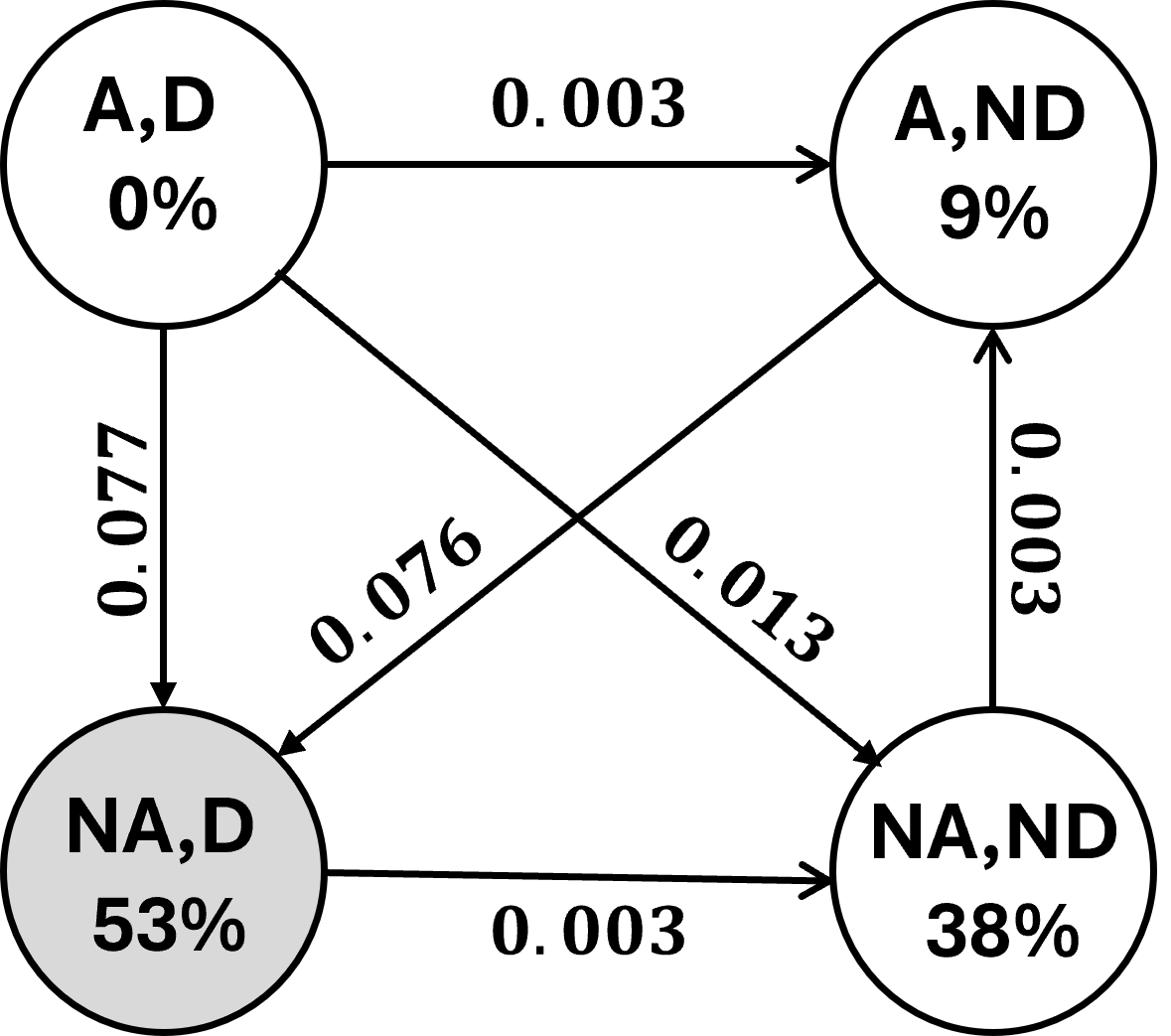}
\caption{}
\end{subfigure}
\hfill
\begin{subfigure}[t]{0.24\textwidth}
\centering
\includegraphics[width=\linewidth]{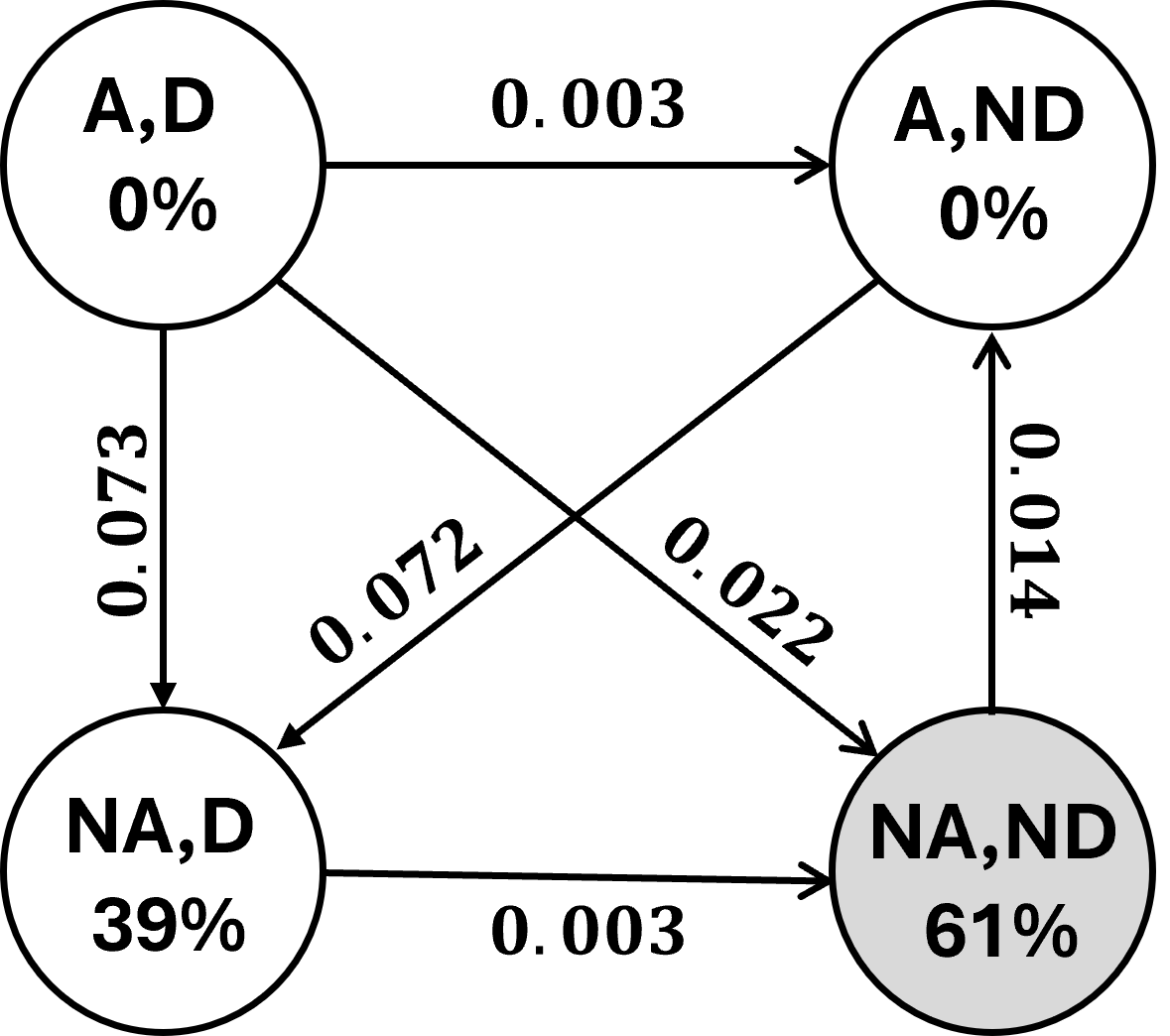}
\caption{}
\end{subfigure}

\caption{
Embedded four-state Markov chains corresponding to the cybersecurity environments shown in Figure~\ref{fig:2} ($\beta=1$). Nodes represent homogeneous population states and are labelled by their stationary probabilities, while directed edges denote fixation probabilities (only the stronger transition between any two states is shown, for clarity). Increasing attack cost and AI-assisted defence effectiveness progressively redirect the evolutionary flow from attack-dominated states towards secure non-attacking strategies. Parameter values for (a) $c_a = 0.2, b_a = 1.0, p_d = 0.2,
c_d = 0.5, b_d = 1.0, w = 1.0$, (b) $c_a = 0.45,b_a = 1.0,p_d = 0.5,c_d = 0.4,b_d = 1.0,w= 1.0$, (c) $c_a = 0.8,b_a = 0.8,p_d = 0.9,c_d = 0.2,b_d = 1.2,w = 1.0$, and (d) $c_a = 0.8,b_a = 0.7,p_d = 0.8;,c_d = 0.2,b_d = 1.2,w = 1.0$.
}
\label{fig:3}
\end{figure}

Panels (a) and (b) represent environments in which AI-assisted attacks remain evolutionarily viable. Under these conditions, stronger selection drives the population towards attack-oriented strategies, with $(A,D)$ dominating when defensive investment remains beneficial and $(A,ND)$ prevailing when maintaining defence becomes comparatively less rewarding. 
Such situations correspond to cyber environments in which AI-assisted offensive tools can be deployed at relatively low cost while defensive mechanisms are insufficiently effective to discourage attacks. 
In contrast, panels (c) and (d) illustrate defence-favourable environments created by higher attack costs and more effective AI-assisted defence. As attacking becomes less profitable, the population  shifts towards non-attacking strategies. The temporary dominance of $(NA,D)$ in panel (c) indicates that defensive behaviour becomes advantageous before attacks are completely eliminated. Under even stronger deterrence, shown in panel (d), the system ultimately converges to $(NA,ND)$, where attacks are no longer evolutionarily attractive and defensive investment is also no longer required.

These transitions directly follow from the incentive structure of the proposed game. Increasing the attack cost reduces the expected payoff of cyber attackers, and improvements in AI-assisted defence decrease the probability of successful attacks. Together, these mechanisms reduce the evolutionary advantage of attacking strategies and progressively shift the population towards secure non-attacking behaviour. In attack-favourable environments, transitions are concentrated towards $(A,D)$ and $(A,ND)$, indicating that these strategies possess the highest fixation probabilities and therefore dominate the long-run dynamics. As attack costs increase and AI-assisted defence becomes more effective, the transition structure shifts towards $(NA,D)$ and eventually $(NA,ND)$ reflects the reduced evolutionary viability of offensive behaviour.

The stationary strategy frequencies are further explained by the embedded Markov chains shown in Figure~\ref{fig:3}. Whereas Figure~\ref{fig:2} describes the final evolutionary outcome, the Markov diagrams reveal how fixation events drive transitions between homogeneous population states. Together, Figures~\ref{fig:2} and~\ref{fig:3} demonstrate that long-run cybersecurity outcomes are governed by the interaction between attack incentives and AI-assisted defensive capability. Improvements in defensive effectiveness not only reduce the success of individual attacks but also alter the overall evolutionary flow of the population towards increasingly secure states.

\begin{figure}[t!]
\centering
    \includegraphics[width=1\textwidth]{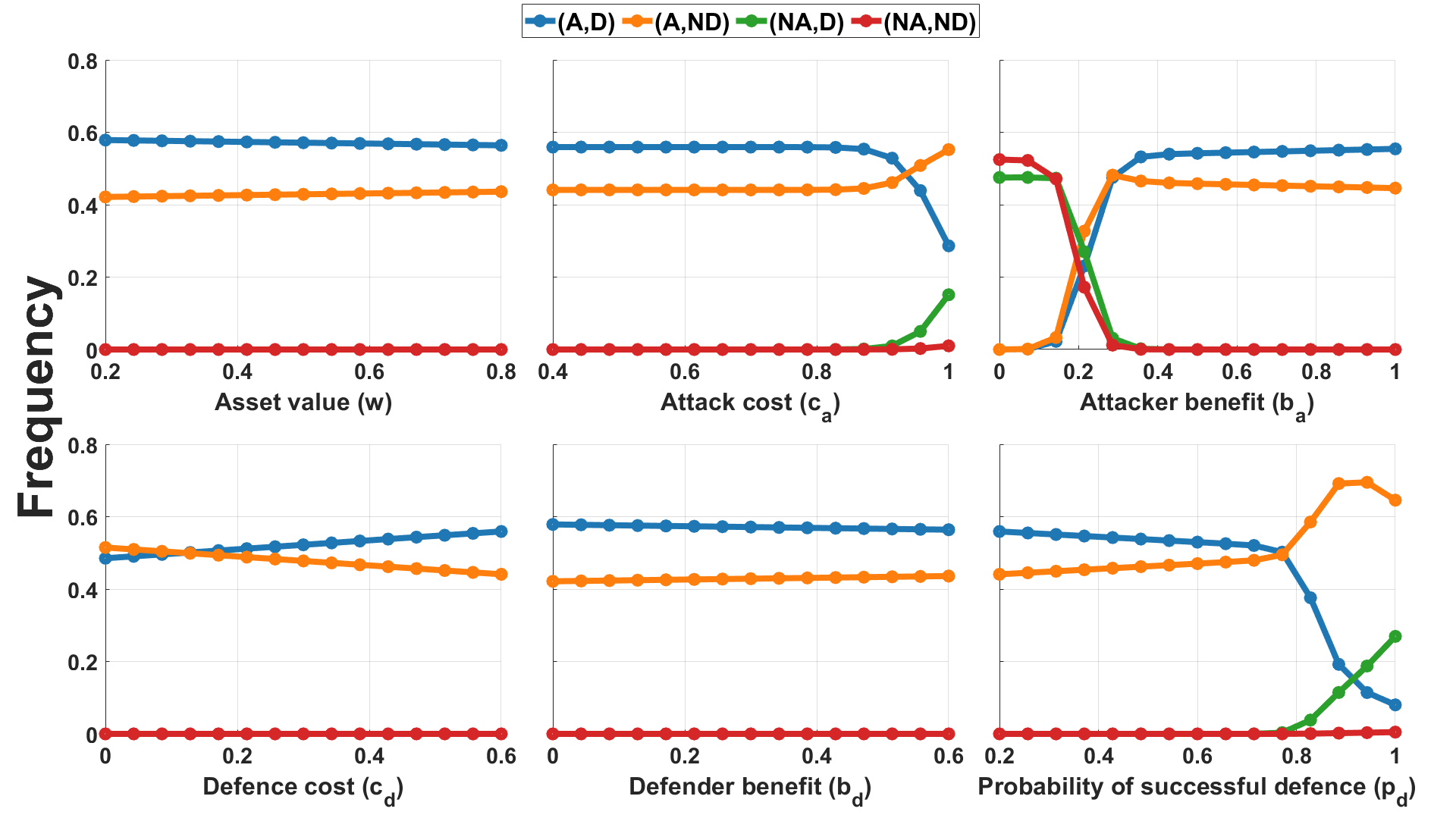}
\caption{
Parameter sensitivity analysis of the stochastic AI-assisted cyber attack-defence model. Each parameter is varied independently over the ranges $w,c_a,c_d,b_d,p_d\in[0,1]$ and $b_a\in[0,2]$, while all remaining parameters are $c_a = 0.2, b_a = 1.0, p_d = 0.2,
c_d = 0.5, b_d = 1.0, w = 1.0$. Results are obtained from the finite-population Markov model with $\beta=1$. Attack cost $c_a$, attacker benefit $b_a$, and AI-assisted defence effectiveness $p_d$ have the strongest influence on long-run evolutionary outcomes, whereas variations in asset value, defence cost, and defender benefit produce comparatively smaller changes.
}
\label{fig:4}
\end{figure}

\subsection{Parameter Sensitivity Analysis}

The previous analysis demonstrated how selection intensity influences the long-run evolution of AI-assisted cyber attack-defence strategies. We next investigate how variations in the underlying cybersecurity parameters affect these evolutionary outcomes. Figure~\ref{fig:4} presents the stationary strategy frequencies obtained from the stochastic Markov model as each parameter is varied independently. Although all parameters influence the evolutionary dynamics, their effects differ considerably. In particular, attack cost $c_a$, attacker benefit $b_a$, and AI-assisted defence effectiveness $p_d$ produce the largest changes in strategy frequencies, whereas variations in asset value $w$, defence cost $c_d$, and defender benefit $b_d$ lead to comparatively smaller changes. These trends directly reflect the evolutionary incentives of the proposed game. Increasing the attack cost reduces the payoff for attackers, making AI-assisted attacks progressively less attractive. Conversely, larger attacker benefits increase the evolutionary advantage of attacking strategies by raising the expected reward from successful attacks. Similarly, improving AI-assisted defence effectiveness lowers the probability of successful attacks and shifts the evolutionary balance towards secure defensive behaviour. These observations are consistent with the analytical attack-suppression condition derived in Section~\ref{sec:model}, which predicts that attacks become evolutionarily unattractive once their expected payoff falls below their operational cost.

From a cybersecurity perspective, these results indicate that increasing the operational cost of attacks and improving the effectiveness of AI-assisted defence are considerably more effective than modifying defensive costs or the value of protected assets alone. Practical mechanisms such as AI-assisted intrusion detection, automated response systems, and exploit mitigation can therefore have a greater long-term impact on suppressing adversarial behaviour than policies that simply increase defensive investment. We next investigate how these evolutionary transitions are distributed across the joint attack cost-defence effectiveness parameter space in well-mixed and structured populations.

\subsection{Simulation Setup}

To validate the stochastic analysis, we perform agent-based simulations for both well-mixed and structured populations using the mixed-role game introduced in Section~\ref{sec:model}. Unless otherwise stated, the mutation probability is fixed at $\mu=10^{-5}$ and the selection intensity at $\beta=0.1$. The remaining model parameters are specified in the corresponding experiments. For the well-mixed population, we consider a finite population of $M=100$ agents. Simulations are performed for $30,000$ update steps, and all reported quantities are averaged over $50$ independent simulation runs to reduce stochastic fluctuations.

For the structured population, agents occupy the nodes of a $100\times100$ square lattice with periodic boundary conditions. Each agent interacts only with its four nearest neighbours, and strategies are updated asynchronously using the Fermi imitation rule described in Eq.~(\ref{eq:2}). One Monte Carlo step (MCS) consists of $L^2$ elementary updates. Each simulation is executed for $30,000$ MCSs, and stationary frequencies are obtained by averaging over the final $5000$ MCSs. To improve statistical reliability, all results are further averaged over $10$ independent runs.

\subsection{Evolutionary Dynamics in Well-Mixed Populations}

\begin{figure}[t!]
\centering
    \includegraphics[width=1\textwidth]{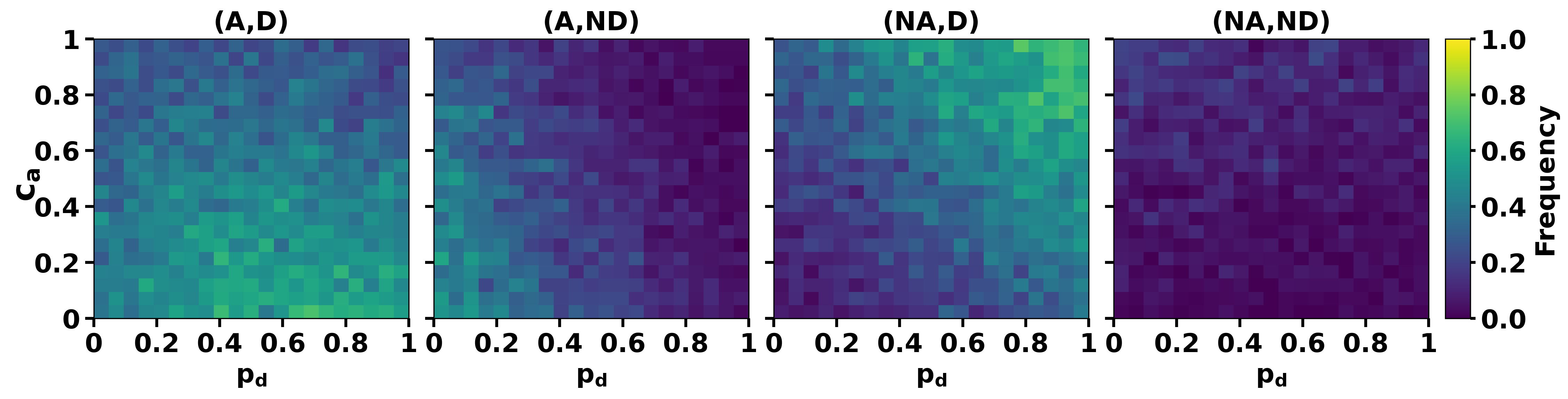}
\caption{
Heatmaps of stationary strategy frequencies in the well-mixed AI-assisted cyber attack-defence model as attack cost $c_a$ and AI-assisted defence effectiveness $p_d$ vary. Other parameters are fixed at $w=1.0$, $b_a=1.0$, $c_d=0.4$, $b_d=1.0$, and $\beta=0.1$. Low attack costs and weak defence favour attack-oriented strategies, whereas increasing either $c_a$ or $p_d$ progressively suppresses attacks and promotes the secure defensive strategy $(NA,D)$. The broad coexistence regions and smooth transitions reflect the stochastic nature of well-mixed evolutionary dynamics.
}
\label{fig:5}
\end{figure}

The parameter sensitivity analysis identified attack cost $c_a$ and AI-assisted defence effectiveness $p_d$ as the dominant factors governing long-run cybersecurity dynamics. We now examine their combined influence across the entire parameter space. Figure~\ref{fig:5} presents heatmaps of the stationary frequencies of the four mixed-role strategies as both $c_a$ and $p_d$ vary simultaneously.
The well-mixed population exhibits broad regions in which multiple strategies coexist, indicating that AI-assisted cybersecurity dynamics remain highly stochastic when interactions occur globally. This behaviour is consistent with the analytical attack-suppression condition derived in Section \ref{sec:model}. At low $c_a$ and $p_d$, attacking strategies dominate because offensive actions remain profitable and defensive protection provides only limited resistance. As either $c_a$ or $p_d$ increases, attacking behaviour gradually declines and the secure defensive strategy $(NA,D)$ expands over a large region of the parameter space.

The figure illustrates that the transitions between strategies are smooth and this behaviour arises because every agent interacts equally with every other agent. Consequently, attacking and defensive behaviours continue to coexist over a wide range of parameter values, producing gradual evolutionary transitions instead of clear strategic boundaries.
These results demonstrate that AI-assisted defence alone can substantially suppress attacks by reducing their evolutionary advantage. However, in a well-mixed population the absence of local organisation limits the ability of defensive behaviour to completely eliminate attacking strategies. This suggests that defensive capability alone may not be sufficient to maximise long-term cyber resilience when interactions occur globally.

The well-mixed population analysis shows that improving $p_d$ and increasing $c_a$ reduce the long-run prevalence of cyber attacks. However, the global interaction pattern also produces broad coexistence regions in which attacking and defensive strategies persist together. These results provide a useful baseline for assessing whether local interactions can further enhance attack suppression and promote stable defensive behaviour.

\subsection{Evolutionary Behaviour in Structured Populations}

Real-world AI-assisted cyber systems are typically organised as interconnected networks rather than homogeneous populations. Organisations, autonomous security agents, and digital devices primarily interact with neighbouring entities through communication links and shared infrastructures. Such local interactions create opportunities for defensive clustering and network reciprocity, which may fundamentally alter the long-run evolutionary dynamics. We therefore investigate how population structure and local interactions influence the evolutionary dynamics of the proposed model.

\begin{figure}[t!]
\centering
    \includegraphics[width=1\textwidth]{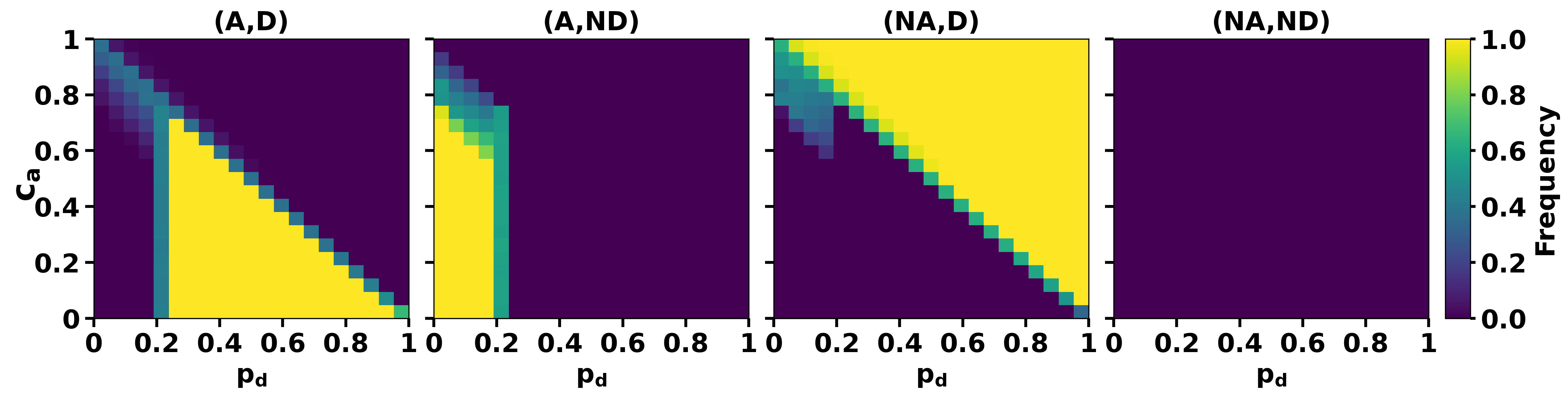}
\caption{
Heatmaps of stationary strategy frequencies in the structured AI-assisted cyber attack-defence model as attack cost $c_a$ and AI-assisted defence effectiveness $p_d$ vary. Other parameters are fixed at $w=1.0$, $b_a=1.0$, $c_d=0.4$, $b_d=1.0$, $\mu=10^{-5}$, and $\beta=0.1$. Compared with the well-mixed population, local interactions produce well-defined strategic regions with sharp evolutionary transitions. Increasing either attack cost or defence effectiveness rapidly promotes the secure defensive strategy $(NA,D)$, demonstrating that population structure substantially enhances long-run cyber resilience.
}
\label{fig:6}
\end{figure}

Figure~\ref{fig:6} presents the stationary strategy frequencies on the square lattice as $c_a$ and $p_d$ vary simultaneously. Compared with the well-mixed population (see again Figure \ref{fig:5}), the structured population exhibits a markedly different evolutionary landscape. Instead of broad coexistence regions, the lattice develops well-defined strategic domains separated by  sharp evolutionary transitions.

The secure defensive strategy $(NA,D)$ occupies a substantially larger region of the parameter space than in the well-mixed population. As either $c_a$ or $p_d$ increases, attacking strategies become less frequent and the population converges towards defence-oriented behaviour. In contrast, the attack-oriented strategy $(A,ND)$ remains viable only when attacks are inexpensive and AI-assisted defence is weak. The $(A,D)$ strategy forms an intermediate transition region between attack-dominated and defence-dominated regimes, whereas $(NA,ND)$ remains negligible throughout the parameter space. Population structure does not alter the boundary condition itself (when a strategy is dominance) but changes how rapidly the population reaches the defence-dominated regime predicted by the condition through network reciprocity. 

These behavioural transitions arise from the local interaction structure of the lattice. Unlike the well-mixed population, neighbouring AI-assisted defenders repeatedly interact with one another and accumulate mutual defensive benefits over time. Once small defensive clusters emerge, they become increasingly resistant to invasion by attacking strategies and gradually expand through local imitation. Consequently, defensive behaviour spreads collectively across the network, leading to much sharper transitions between attack-dominated and defence-dominated regimes than those observed under global mixing.

From a cybersecurity perspective, these findings suggest that interaction structure itself contributes to cyber resilience. Organising AI-assisted defence through locally connected infrastructures enables neighbouring defensive agents to reinforce one another, allowing secure behaviour to emerge under conditions where globally mixed populations continue to exhibit persistent attacks. Although the lattice heatmaps clearly demonstrate that population structure favours secure defensive behaviour, they do not explain how this transition occurs over time. We therefore examine the spatial evolution of strategies to identify the mechanism responsible for attack suppression and the emergence of stable defensive regions.

\subsection{Defensive Clustering as a Mechanism for Attack Suppression}

The lattice heatmaps, in Figure~\ref{fig:6}, reveal that local interactions substantially enlarge the region in which secure defensive behaviour dominates. To understand the underlying mechanism, we next examine the spatial evolution of strategies under different cybersecurity conditions. The spatial snapshots illustrate how local interactions reorganise the population over time and reveal the process through which AI-assisted defence suppresses persistent attacks.

\begin{figure}[t]
\centering
    \includegraphics[width=0.70\textwidth]{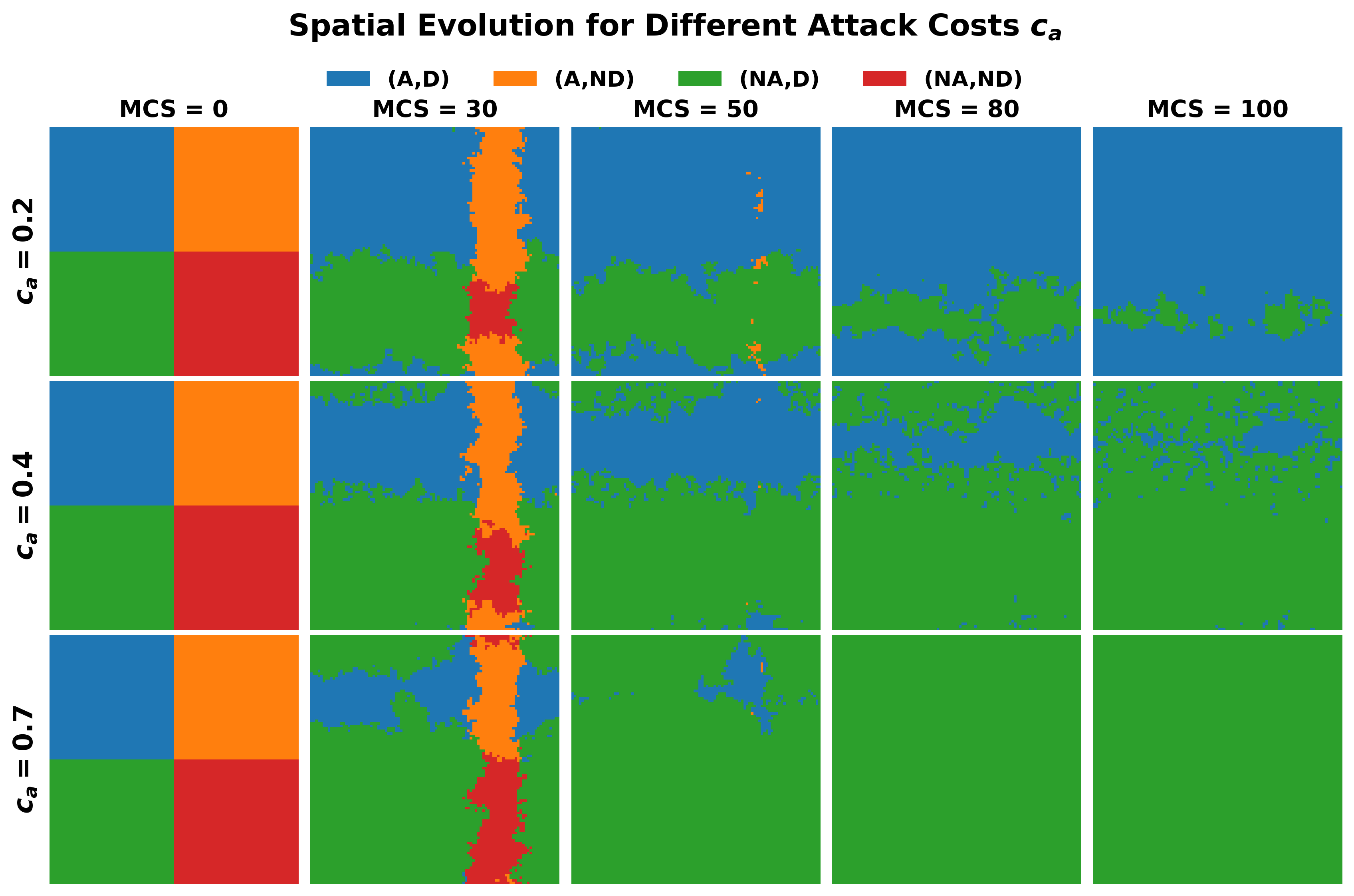}
\caption{
Spatial evolution of AI-assisted cyber attack-defence strategies for different attack costs $c_a$. Columns correspond to $MCS=\{0,30,50,80,100\}$ and rows to $c_a=\{0.2,0.4,0.7\}$. Other parameters are fixed at $w=1.0$, $b_a=0.7$, $c_d=0.3$, $b_d=0.8$, $p_d=0.5$, and $\beta=0.1$. Increasing the attack cost progressively fragments attacking clusters and promotes the expansion of the secure defensive strategy $(NA,D)$ through local interactions.
}
\label{fig:7}
\end{figure}

Figure~\ref{fig:7} investigates the influence of attack cost on the spatial evolution of the population. Initially, all four strategies are randomly distributed across the lattice. When attacks remain inexpensive $(c_a=0.2)$, attack-oriented strategies persist throughout the network because offensive actions continue to generate sufficient evolutionary rewards. As the attack cost increases, attacking clusters progressively fragment and shrink, while neighbouring defensive agents expand into the vacated regions. At sufficiently high $c_a$, the secure defensive strategy $(NA,D)$ occupies almost the entire population, indicating that persistent attacks are no longer evolutionarily sustainable.

The underlying mechanism is local reinforcement. Because neighbouring defenders repeatedly interact with one another, they accumulate higher collective payoffs than isolated attackers. Once a defensive cluster becomes sufficiently large, it becomes increasingly difficult for attacking strategies to invade. Instead, attackers survive only along cluster boundaries, where they are gradually eliminated through repeated imitation of more successful neighbouring defenders. Consequently, increasing $c_a$ can accelerate the expansion of defensive clusters and suppress the long-run persistence of attacks. We next investigate whether strengthening AI-assisted defence $p_d$ produces a similar evolutionary mechanism.

\begin{figure}[t!]
\centering
    \includegraphics[width=0.70\textwidth]{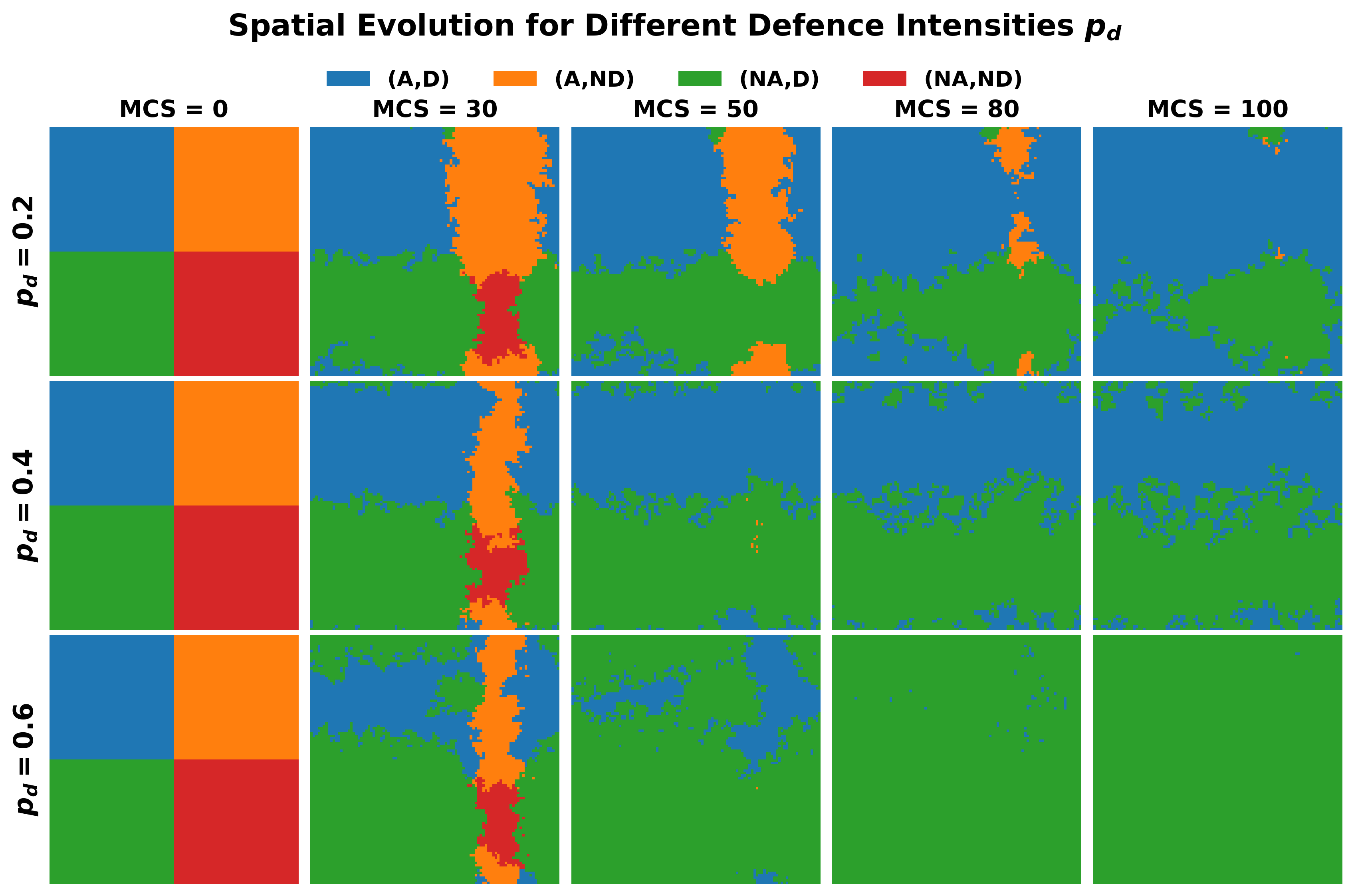}
\caption{
Spatial evolution of AI-assisted cyber attack--defence strategies for different AI-assisted defence effectiveness $p_d$. Columns correspond to $MCS=\{0,30,50,80,100\}$ and rows to $p_d=\{0.2,0.4,0.6\}$. Other parameters are fixed at $w=1.0$, $b_a=0.7$, $c_a=0.4$, $c_d=0.3$, $b_d=0.8$, and $\beta=0.1$. Stronger AI-assisted defence accelerates the formation and expansion of defensive clusters, rapidly suppressing persistent attacks and promoting the secure defensive strategy $(NA,D)$.
}
\label{fig:8}
\end{figure}

Figure~\ref{fig:8} presents the spatial evolution obtained under different levels of AI-assisted defence effectiveness. When defensive capability is weak, attacking strategies remain active for extended periods because successful attacks continue to provide sufficient evolutionary rewards. As defence effectiveness increases, attacking clusters contract more rapidly and are progressively replaced by expanding defensive regions. Under strong AI-assisted defence, the population quickly converges towards the secure defensive strategy $(NA,D)$.

Although $c_a$ and $p_d$ influence different components of the payoff structure, both parameters alter the same evolutionary balance between attack and defence. Increasing attack cost discourages offensive behaviour directly, whereas improving AI-assisted defence reduces the probability of successful attacks. In both cases, defensive clusters obtain higher cumulative payoffs than neighbouring attackers, allowing secure behaviour to propagate across the lattice through local imitation.

These findings have important implications for AI-assisted cyber defence. Modern cybersecurity platforms increasingly deploy distributed autonomous defence agents across enterprise networks, cloud infrastructures, industrial control systems, and edge devices. Our results suggest that when these agents interact primarily with neighbouring systems, local coordination naturally strengthens collective defence. Even without introducing additional incentives, local defensive organisation substantially improves the ability of the network to suppress persistent attacks.

\section{Conclusion}
\label{sec:conclusion}

AI-assisted cybersecurity systems are characterised by continuous adaptation between attackers and defenders, making EGT a suitable framework for understanding their long-term behaviour. However, existing evolutionary cybersecurity models have primarily focused on homogeneous interactions, providing limited insight into how network structure influences the evolution of attack and defence. Understanding this relationship is becoming increasingly important as modern cyber systems rely on interconnected AI-enabled agents operating across enterprise networks, cloud infrastructures, industrial control systems, and IoT environments.

To address this gap, we developed a mixed-role evolutionary game for AI-assisted cyber attack-defence interactions and investigated its dynamics in both well-mixed and structured populations. By combining stochastic evolutionary analysis with large-scale agent-based simulations, we examined how interaction structure influences the long-run balance between attacking and defensive behaviour. Unlike conventional attacker-defender models with fixed roles, the proposed framework allows agents to exhibit both offensive and defensive behaviours, providing a more flexible representation of modern AI-assisted cybersecurity environments.

Our findings demonstrate that population structure fundamentally changes cybersecurity behavioural evolution. While well-mixed populations exhibit broad coexistence between attacking and defensive strategies, structured populations generate well-defined evolutionary regimes in which the secure defensive strategy $(NA,D)$ occupies a substantially larger region of the parameter space. The spatial analysis further reveals that local interactions enable neighbouring defenders to form stable defensive clusters that progressively suppress attacking behaviour through network reciprocity. These observations are consistent with previous studies demonstrating the importance of interaction structure in evolutionary dynamics \cite{szabo2007evolutionary,perc2017statistical,cimpeanu2022artificial,pi2026evolutionary}, while extending these concepts to AI-assisted cyber attack-defence systems.

From a practical cybersecurity perspective, the results suggest that improving cyber resilience requires more than increasing attack costs or deploying stronger defensive technologies. The organisation of interactions between AI-assisted defensive agents also plays an important role. Designing cyber systems that encourage local coordination and collaboration among neighbouring defenders can naturally strengthen collective defence and reduce the long-term persistence of attacks.

The present work also opens several directions for future research. More realistic network topologies, including scale-free, small-world, and random networks, should be investigated to evaluate the robustness of the proposed framework \cite{szabo2007evolutionary}. The model can also be extended to incorporate heterogeneous AI agents, adaptive learning, reputation mechanisms, cyber threat intelligence sharing, and multi-stage attack scenarios. These extensions will further improve the realism of evolutionary cybersecurity models and contribute towards the design of more resilient, network-aware AI-assisted cyber ecosystems.

\section*{Acknowledgements}
T.A.H. and Z.S. are supported by EPSRC (grant EP/Y00857X/1).

\bibliographystyle{unsrt}
\bibliography{Structured}

\end{document}